\documentclass[showpacs,preprint]{revtex4}

\usepackage{graphicx}
\usepackage{dcolumn}
\usepackage{amsmath}

\begin{document}


\title{Exact barotropic distributions in Einstein-Gauss-Bonnet gravity}


\author{Sunil D. Maharaj}
\author{Brian Chilambwe}
\author{Sudan Hansraj}
\altaffiliation{}
\affiliation{Astrophysics and Cosmology Research Unit, School of Mathematics, Statistics and Computer Science, University of KwaZulu--Natal, Private Bag X54001, Durban 4000, South Africa}
\email[]{maharaj@ukzn.ac.za}
\email[]{brian@aims.ac.za}
\email[]{hansrajs@ukzn.ac.za}

\date{\today}

\begin{abstract}

New exact solutions to the field equations in the Einstein--Gauss--Bonnet modified theory of gravity for a 5--dimensional spherically symmetric static distribution of a perfect fluid is obtained. The Frobenius method is used to obtain this solution in terms of an infinite series. Exact solutions are generated in terms
of polynomials from the infinite series.
The 5--dimensional Einstein solution is also found by setting the coupling constant to be zero. All models admit a barotropic equation of state. Linear equations of state are admitted in particular models with the energy density profile of isothermal distributions. We examine the physicality of the solution by studying graphically the isotropic pressure and the energy density. The model is well behaved in the interior and the weak, strong and dominant energy conditions are satisfied.

\end{abstract}


\maketitle


\section{Introduction}

In many respects the general theory of relativity proposed by Einstein continues to be the most successful theory of the gravitational field. However it does come short in explaining certain observed phenomena. For example, the late time expansion of the universe is not a direct consequence of the standard Einstein theory but is reported in experimental observations. One possible approach to correct this deficiency in the Einstein gravity is to allow the action principle to include more than just linear forms of the Riemann tensor, the Ricci tensor and the Ricci scalar. The choice of just linear tensorial quantities has the advantage of producing second order equations of motion which are compatible with the standard Einstein theory in four dimensions.  Lovelock \cite{1,2} proposed a polynomial form of the Lagrangian;  if this is taken to be of quadratic order we generate the Einstein--Gauss--Bonnet (EGB) action. The amazing feature associated with the EGB Lagrangian is that the equations of motion continue to be second order quasi--linear. If the higher order effects are absent then, the regular Einstein field equations are regained \cite{3}. Thus far, researchers in the field have reported numerous results involving exterior solutions in 5--dimensional EGB gravity theory. For example Anabalon et al \cite{4} found an exact vacuum solution in five dimensions with the Kerr-Schild ansatz in EGB gravity, and the vacuum Boulware--Deser \cite{5}  exterior solution is a well known model. Issues related to gravitational collapse have received much attention as well. For instance, the case of collapsing dust with zero pressure in 5--dimensional EGB theory has been well studied by Maeda \cite{6}. 

The model that we study in this paper turns out to have the density profile of an isothermal sphere. Isothermal spheres have the energy density being inversely proportional to the the square of their radius. These spheres have been widely studied due to their importance as models for different astronomical objects such as globular and open clusters, galactic bulges, elliptical galaxies and clusters of galaxies as indicated by Milgrom \cite{7}. Saslaw et al \cite{8} considered the role of isothermal spheres in inhomogeneous cosmological models in general relativity. Some other past works involving isothermal spheres are gravitational instabilities in the presence of a cosmological constant \cite{9}, gravitational collapse \cite{10, 11, 12} and gravitational lensing properties of isothermal  spheres with a finite core \cite{13} and thermodynamics in bounded self-gravitating isothermal spheres \cite{14}. It is interesting that isothermal density profiles also arise in EGB theory in five dimensions as we will demonstrate.

In this paper we seek new exact interior models in 5--dimensional  EGB  theory with a spherical distribution of perfect fluid. Some attempts in this direction have been made by Kang et al \cite{15} and Dadhich et al \cite{16}, in EGB gravity without a cosmological constant. The first gravitational potential is specified. We then express the EGB field equations in standard canonical coordinates, and then introduce a coordinate transformation which allows the single master field equation to be written as a second order ordinary differential equation in the remaining gravitational potential. Our approach allows us to solve the master differential equation in closed form. To integrate the master field equation we utilise the method of Frobenius. Solutions are possible in terms of series and polynomials. Our models are characterised by a barotropic equation of state.

In Section II we briefly discuss the basic principles of EGB theory. The EGB field equations, used to describe gravitational behaviour of 5--dimensional EGB gravity in static spherical fields, are derived in Section III. In Section IV we present new classes of exact interior solutions. These are valid for both 5--dimensional EGB theory and the 5--dimensional Einstein gravity cases. Elementary functions which arise from the general solution in Section IV are presented in Section V. The Physical properties of the model are examined in Section VI. To verify the physical reasonableness of the model we perform a graphical analysis. We make our conclusions in Section VII.

\section{Einstein--Gauss--Bonnet Gravity}

We require an action to generate the field equations in EGB gravity. In this paper we are interested in five dimensions. The Gauss--Bonnet action in five dimensions has the form
\begin{equation}
S = \int \sqrt{-g} \left[ \frac{1}{2} \left(R - 2\Lambda + \alpha L_{GB}\right)\right] d^5 x + S_{\mbox{ matter}}, \label{1}
\end{equation}
where the parameter $ \alpha $ represents the Gauss--Bonnet coupling constant. We observe that the Lagrangian is quadratic in the Ricci tensor, Ricci scalar  and the Riemann tensor. However the advantage of this action is that the equations of motion turn out to be second order quasilinear which is a distinguishing feature. The Gauss--Bonnet term makes no contribution for $ n \leq 4 $ but has a nonzero value for $ n > 4 $.

The EGB field equations can be expressed in the form
\begin{equation}
G_{a b} + \alpha H_{a b} = T_{a b},  \label{2}
\end{equation}
with metric signature $ (- + + + +) $. The quantity $ G_{ab} $ is the Einstein tensor. The Lanczos tensor $ H_{a b} $ is defined by
\begin{equation}
H_{a b} = 2 \left(R R_{a b} - 2 R_{a c}R^{c}_{b} - 2 R^{c d} R_{a c b d} + R^{c d e}_{a} R_{b c d e} \right) - \frac{1}{2} g_{a b} L_{G B}.  \label{3}
\end{equation}
 The Lovelock term is given by
\begin{equation}
L_{G B} = R^2 + R_{a b c d} R^{a b c d} - 4R_{c d} R^{c d},   \label{4}
\end{equation}
which is a specific combination of Ricci scalar, the Ricci tensor and the Riemann tensor. 
\section{Field Equations}
As we are concerned with five dimensions we take the line element for static spherically symmetric spacetimes to be of the form
\begin{equation}
ds^{2} = -e^{2 \nu} dt^{2} + e^{2 \lambda} dr^{2} + r^{2} \left( d\theta^{2} + \sin^{2} \theta d \phi^{2} + \sin^{2} \theta \sin^{2} \phi d\psi^{2} \right), \label{5}
\end{equation}
where $ \nu(r) $ and $ \lambda(r) $ are arbitrary functions representing the gravitational field. We use a comoving fluid velocity $ u^a = e^{-\nu} \delta_{0}^{a} $ which is timelike and unit. The matter field is described by a perfect fluid with energy momentum tensor of the form
\begin{equation}
T_{a b} = (\rho + p) u_a u_b + p g_{a b}, \label{6} 
\end{equation}
where $ \rho $ and $ p $ are the energy density and isotropic pressure respectively.

 Then the EGB field equations (\ref{2}) may be expressed as
\begin{eqnarray}
\rho &=& \frac{3}{e^{4 \lambda} r^{3}} \left( r e^{4 \lambda} - r e^{2 \lambda} - 4 \alpha \lambda ^{\prime} + r^{2} e^{2 \lambda} \lambda ^{\prime} + 4 \alpha e^{2 \lambda} \lambda ^{\prime} \right),  \label{7a} \\ \nonumber \\
p &=&  \frac{3}{e^{4 \lambda} r^{3}} \left(-  r e^{4 \lambda} + \left( r^{2} \nu^{\prime} +  r + 4 \alpha \nu^{\prime} \right) e^{2 \lambda} - 3 \alpha \nu^{\prime} \right),  \label{7b} \\ \nonumber \\
p &=& \frac{1}{e^{4 \lambda} r^{2}} \left( -e^{4 \lambda} - 4 \alpha \nu^{\prime \prime} + 12 \alpha \nu^{\prime} \lambda^{\prime} - 4 \alpha \left( \nu^{\prime} \right)^{2}  \right) \nonumber \\
                 & \quad & + \frac{1}{e^{2 \lambda} r^{2}} \left(  1 - r^{2} \nu^{\prime} \lambda^{\prime} + 2 r \nu^{\prime} - 2 r \lambda^{\prime} + r^{2} \left( \nu^{\prime} \right)^{2}  \right) \nonumber \\
                 & \quad & + \frac{1}{e^{2 \lambda} r^{2}} \left(  r^{2} \nu^{\prime \prime} - 4 \alpha \nu^{\prime} \lambda^{\prime} + 4 \alpha \left( \nu^{\prime} \right) ^{2} + 4 \alpha \nu^{\prime \prime}   \right). \label{7c}
\end{eqnarray}
The field equations are highly nonlinear; the appearance of terms associated with $ \alpha $ leads to additional complexity. The system (\ref{7a})--(\ref{7c}) consists of three field equations in four unknowns which is similar to the four dimensional field equations in the Einstein limit for spherically symmetric perfect fluids. We regain the Einstein limiting case when $ \alpha = 0 $. 

The transformation 
\begin{equation}
e^{2 \nu} = y^{2}(x), \, \, e^{-2 \lambda} = Z(x), \, \, x = r^{2}, \label{8}
\end{equation} 
was introduced by Durgapal and Bannerji \cite{17}. This transformation has been successfully utilised in the Einstein case to generate exact solutions. Examples of simple metrics are provided by Finch and Skea \cite{18} and Hansraj and Maharaj \cite{19} for neutral and charged isotropic spheres respectively. For recent examples to charged anisotropic relativistic stellar models, found with the help of this transformation, see the models of Mafa Takisa and Maharaj \cite{20}, Maharaj et al \cite{21} and Thirukkanesh and Maharaj \cite{22} in four dimensions. We apply the transformation (\ref{8}) to the 5--dimensional EGB equations. Then the field equations (\ref{7a})--(\ref{7c}) may be written as
\begin{eqnarray}
\frac{3  (1 - Z) ( 1 - 4 \alpha \dot{Z} )}{x} - 3  \dot{Z} &=& \rho, \label{9a} \\ \nonumber \\
\frac{6  Z \dot{y}}{y} + \frac{24 \alpha (1 - Z) Z \dot{y}}{x y} - \frac{3  (1 - Z)}{x} &=& p, \label{9b} \\ \nonumber \\
 2 x Z \left( 4 \alpha [Z - 1] - x \right) \ddot{y}  - \left( x^{2} \dot{Z} + 4 \alpha \left[ x \dot{Z} - 2 Z + 2 Z^{2} - 3 x Z \dot{Z} \right] \right) \dot{y} \nonumber \\ - \left( 1 + x \dot{Z} - Z \right) y &=& 0.  \label{9c}
\end{eqnarray}
The last equation is the generalisation of the equation of pressure isotropy. Equation (\ref{9c}) has been written as a second order differential equation in $ y $; in 4--dimensional Einstein models this proves to be a useful form. We seek exact solutions to the highly nonlinear generalised pressure isotropy condition (\ref{9c}) in the presence of $ \alpha $. When $ \alpha = 0 $ then (\ref{9c}) becomes
\begin{equation}
2 x^{2} Z \ddot{y} + x^{2} \dot{Z} \dot{y} + \left( 1 + x \dot{Z} - Z \right) y = 0.  \label{10}
\end{equation}
Clearly (\ref{9c}) is more difficult to integrate in the 5--dimensional EGB case because of its greater complexity and nonlinearity. Note that the 4--dimensional version of the special case (\ref{10}) was comprehensively studied by Thirukkanesh and Maharaj \cite{22}.

\section{New Exact Interior Solutions}

To integrate (\ref{9c}) it is necessary to make simplifying assumptions. Hansraj et al \cite{23} found several classes of exact solutions by essentially choosing forms for the function $ Z $. We observe that if the potential $ Z $ is specified then we can treat (\ref{9c}) as a second order linear differential equation in the function $ y $. This approach may lead to exact solutions in terms of elemenatry functions. We make the simple choice
\begin{equation}
Z = a, \label{11}
\end{equation}
where $ a $ is a constant in our approach. Then equation (\ref{9c}) reduces to
\begin{equation}
x (x + A) \ddot{y} - A \dot{y} + E y = 0, \label{12}
\end{equation}
where we have set
\begin{eqnarray}
A &=& 4 \alpha (1 - a), \label{13a} \\
E &=& \frac{1 - a}{2 a}, \label{13b}
\end{eqnarray}
for convenience. 

Note that if we introduce a new variable $ z = \frac{x + A}{A} $ then (\ref{12}) can be written in the form
\begin{equation}
z (z - 1) \frac{d ^{2} y}{d z ^{2}} - \frac{d y}{d z} + E y = 0, \label{13c}
\end{equation}
which is the hypergeometric differential equation. In general (\ref{13c}) admits solutions in terms of special functions, namely the hypergeometric functions. As equations (\ref{12}) and (\ref{13c}) are equivalent we can integrate either equation. The general solution is expressable in the form of infinite series but polynomial solutions are also permitted.

\subsection{The case $ a = 1 $}

We take $ a = 1. $ Then (\ref{13c}) reduces to
\begin{equation}
x^{2} \ddot{y} = 0, \label{14}
\end{equation}
which is solved to give
\begin{equation}
y = C_{1} + C_{2} x, \label{15}
\end{equation}
\noindent where $ C_{1} $ and $ C_{2} $ are integration constants. From (\ref{9a})  we get the density to be $ \rho = 0 $, and we do not pursue this case further.

\subsection{The case $ a \neq 1 $} 

When $ a \neq 1 $ then (\ref{13c}) is not contained in any of the standard cases of known solutions for differential equations. Consequently we can use the Frobenius method to solve (\ref{13c}) as $ x = 0 $ is a regular singular point. 

Take the first solution to be of the form
\begin{equation}
y_{1}(x) = \sum _{n = 0} ^{\infty} a_{n} x ^{n + c}, \label{16}
\end{equation}
where $ a_{n} $ is the coefficient of the series and $ c $ is a constant. Since the equation (\ref{13c}) is the hypergeometric differential equation we can write the first solution $ y_1(x) $ in the form
\begin{equation}
y_{1}(x) = \sum _{n = 1} ^{\infty} \frac{ 2 a_{0} (-1)^{n}}{A^{n}} \frac{ 1 }{n! (n + 2)!} \prod_{j = 1} ^{n} (j (j + 1) + E) x^{n + 2}, \, \, n \geq 1 \label{17}
\end{equation}
where the symbol $ \prod $ denotes multiplication. In the above $ a_0 $, $ A $ and $ E $ are constants. It is clear that the first solution $ y_1(x) $ is an infinite series.

To find the second linearly independent solution $ y_{2} $ to the differential equation (\ref{13c}) we observe that the roots of the indicial equation differ by an integer. Then the second solution is given by
\begin{equation}
y_{2} (x) = \mu y_{1} (x) \ln x + \sum _{n = 0} ^{\infty} b_{n} x^{n}, \label{18}
\end{equation}
where $ \mu $ is some constant and $ b_{n} $ is the coefficient of the series. For completeness it is necessary that we find a functional form for the coefficient $ b_{n} $. Unfortunately we cannot solve the resulting recurrence relation for $ b_{n} $ for the second solution in general. This contrasts with the first solution $ y_{1} (x) $. However it is always possible to explicitly generate the individual coefficients. The general solution to (\ref{13c}) has the structure
\begin{equation}
y(x) = C_{1} y_{1} (x) + C_{2} y_{2} (x), \label{19}
\end{equation}
where $ y_{1} (x) $ and $ y_{2} (x) $ are given by (\ref{17}) and (\ref{18}) respectively.

\subsection{The case $ \alpha = 0 $}

When $ \alpha = 0 $ in (\ref{12}) we obtain the differential equation
\begin{equation}
x ^{2} \ddot{y} + E y = 0, \label{20}
\end{equation}
which corresponds to the 5--dimensional Einstein case. This has solution
\begin{equation}
y = C_{1} x^{\frac{1 - \sqrt{\frac{3 a - 2}{a}}}{2}} + C_{2} x^{\frac{1 + \sqrt{\frac{3 a - 2}{a}}}{2}}. \label{21}
\end{equation}
The energy density and pressure have the form
\begin{eqnarray}
\rho & = & \frac{3  (1 - a)}{x}, \label{22} \\
p    & = & \frac{3  (a - 1)}{x} \nonumber \\
     & \quad & + \frac{ 3 a \left[ C_{1} \left( 1 - \sqrt{\frac{3 a - 2}{a}} \right) x ^{- \left( \frac{1 + \sqrt{\frac{3 a - 2}{a}}}{2} \right) } + C_{2} \left( 1 + \sqrt{\frac{3 a - 2}{a}} \right) x ^{ - \left( \frac{ 1 - \sqrt{\frac{3 a - 2}{a}}}{2} \right)} \right] }{C_{1} x^{\frac{1 - \sqrt{\frac{3 a - 2}{a}}}{2}} + C_{2} x^{\frac{1 + \sqrt{\frac{3 a - 2}{a}}}{2}}}. \label{23}
\end{eqnarray}
An interesting case arises when we set $ C_{1} = 0 $. Then from (\ref{22}) and (\ref{23}) we get
\begin{equation}
p = \left[ \frac{a}{ (1 - a)} \left( 1 + \sqrt{\frac{3 a - 2}{a}} \right) - 1 \right] \rho, \label{24}
\end{equation}
and we obtain a linear barotropic equation of state in the 5--dimensional Einstein case. In (\ref{24}) we observe that the density profile $ \rho \sim r^{- 2} $ which is of the form for an isothermal sphere. For a discussion of the isothermal inhomogeneous models where pressure balances gravity see the analysis of Saslaw et al \cite{8} in four dimensions.

\section{Elementary Functions}

The first solution $ y_{1} $ in (\ref{17}) is in the form of a series which defines a hypergeometric function. For particular values of the parameter $ E $ it is possible to write the solution in terms of elementary functions which will be polynomials. Such a reduced form is more helpful for discussing the physical features. If we set $ E = - m $, an integer, then the general coefficient can be written in the form 
\begin{equation}
a_{n} = \frac{ 2 a_{0}}{A^{n}} \frac{ 1 }{n! (n + 2)!} \prod_{j = 1} ^{n} ( m- j (j + 1)), \, \, 1 \leq n \leq m. \label{25}
\end{equation}
Then the first solution $ y_{1}(x) $ to the differential equation (\ref{13c}) has the form
\begin{eqnarray}
y_{1}(x) &=& \sum _{n = 1} ^{m} \frac{ 2 a_{0}}{A^{n}} \frac{ 1 }{n! (n + 2)!} \prod_{j = 1} ^{n} ( m- j (j + 1)) x ^{n + 2}. \label{26}
\end{eqnarray}
Therefore the hypergeometric series terminates and the solution is in terms of simple polynomial functions. This behaviour is also exhibited in the neutral and charged stellar models of John and Maharaj \cite{24}, Thirukkanesh and Maharaj \cite{21,25} and Maharaj and Komathiraj \cite{26} in 4--dimensional Einstein models. 

\section{Physical Properties}

The matter distribution should be well behaved. We require that the gravitational potentials and matter variables are regular, causality is maintained and energy conditions should be satisfied in the 5--dimensional EGB spherically symmetric model found in this paper.

Our model has the interesting feature of allowing the barotropic equation of state in general. We observe from (\ref{9a}) that
\begin{equation}
x = 3 C (1 - a) \rho ^{- 1}. \label{27}
\end{equation} 
Since $ Z $ is constant, and the variable $ x $ is expressable in terms of $ \rho $ only the function $ y(x) $ in (\ref{19}) can therefore be written in terms of $ \rho $ only. Thus the pressure $ p $ in (\ref{9b}) can be written in terms of the energy density only:
\begin{equation}
p = p(\rho).  \label{28}
\end{equation}
Thus the model in this paper obeys a barotropic equation of state. A similar situation arises in the charged analogue of Finch--Skea stars in four dimensional general relativity as established by Hansraj and Maharaj \cite{19}.

We now consider the matching at the boundary of the gravitating body. For physical viability, any stellar interior solution should match smoothly to the appropriate exterior spacetime. The junction conditions for matching across stellar surface in EGB gravity is contained in the treatment of Davis \cite{27}. The higher order curvature terms modify the usual junction conditions of conventional Einstein gravity. The junction conditions must be satisfied to demonstrate a complete model. The exterior spacetime is taken to be the Boulware--Deser metric \cite{5}
\begin{equation}
ds^2 = - F(r) dt^2 + \frac{dr^2}{F(r)} + r^{2} \left( d\theta^{2} + \sin^{2} \theta d \phi ^{2} + \sin^{2} \theta \sin^{2} \phi d\psi ^{2} \right), \label{29}
\end{equation}
where
\[
F(r) = 1 + \frac{r^2}{4\alpha} \left( 1 - \sqrt{1 + \frac{8M\alpha}{r^4}} \right),
\]
and $ M $ is the mass of the gravitating hypersphere. For a constant density body in 5--dimensional EGB theory the matching across the boundary was performed by Dadhich et al \cite{16}. In our variable density model we match the metrics (\ref{5}) and (\ref{29}) across the boundary $ r = R $. This yields the conditions
\begin{eqnarray}
1 + \frac{R^2}{4\alpha} \left( 1 - \sqrt{1 + \frac{8M \alpha}{R^4}} \right) &=& y^{2}(R^{2}) = e^{2 \nu(R)} \label{30} \\
1 + \frac{R^2}{4\alpha} \left( 1 - \sqrt{1 + \frac{8M\alpha}{R^4}} \right) &=& Z(R^{2}) = e^{-2 \lambda(R)} \label{31}
\end{eqnarray}
With the help of (\ref{9b}), (\ref{11}) and (\ref{19}), the vanishing of the pressure at the boundary requires
\begin{equation}
2 a \left( R^{2} + 4 \alpha (1 - a) \right) (C_1 \dot{y}_1 + C_2 \dot{y}_2) - (1 - a) (C_1 y_1 + C_2 y_2)  = 0. \label{32}
\end{equation}
Then we can show that (\ref{30})--(\ref{32}) admits the solution
\begin{subequations}
\label{33}
\begin{eqnarray}
C_1 &=& \frac{\sqrt{a} - C_2 y_2}{y_1}, \label{33a} \\
C_2 &=& \frac{ \left[ (1 - a) y_1 - 2 a (R^2 + 4 \alpha (1 - a)) \dot{y}_1 \right] }{ 2 \sqrt{a} \left[ R^2 + 4 \alpha (1 - a) \right] (y_1 \dot{y}_2 - y_2 \dot{y}_1)}, \label{33b} \\
a   &=& 1 + \frac{R^2}{4\alpha} \left( 1 - \sqrt{1 + \frac{8M\alpha}{R^4}} \right). \label{33c}
\end{eqnarray}
\end{subequations}
This uniquely fixes the arbitrary constants $ a $, $ C_1 $ and $ C_2 $ in terms of the stellar radius $ R $, the mass $ M $ and the Gauss--Bonnet parameter $ \alpha $.

Now to study the physical features. For the physical analysis we truncate (\ref{17}) to get the first three terms as
\begin{equation}
y = -\frac{2 (2 + E) a_{0}}{ 1! 3! A} x^3 + \frac{2 (2 + E) (6 + E) a_{0}}{ 2! 4! A^{2}} x^4 - \frac{2 (2 + E) (6 + E) (12 + E) a_{0}}{ 3! 5! A^{3}} x^5. \label{34}
\end{equation}
This is always possible in general since we showed in Section V that polynomial solutions are permitted in the first solution $ y_{1} (x) $. Using (\ref{9a}) we find the density to be
\begin{equation}
\rho = \frac{3 (1 - a)}{x}, \label{35}
\end{equation}
From (\ref{9b}) the pressure is given by
\begin{equation}
p = \frac{3 (a - 1)}{x} + 6 a \left( \frac{x - 4 \alpha (a - 1)}{x^{2}}\right) K(x), \label{36}
\end{equation}
where
\begin{equation}
K(x) =  \left[ \frac{360 A^{2} - 60 A (6 + E ) x + 5 (6 + E) (12 + E) x^{2} }{120 A^{2} - 15 A (6 + E) x + (6 + E) (12 + E) x^{2} } \right].
\end{equation}
Plots for the energy density $ \rho $ and pressure are given in Fig. 1-2 respectively. We note that they are both decreasing functions. 

As $ \rho $ and $ p $ are known explicitly we can calculate the quantity $ \frac{d p}{d \rho} $. This is given by
\begin{equation}
\frac{d p}{d \rho} = \frac{2 a \left[ \left\lbrace x - 2 (x - 4 \alpha (a - 1)) \right\rbrace K (x) + x (x - 4 \alpha (a - 1)) K^{\prime} (x) \right]}{(1 - a) x} - 1, \label{37} 
\end{equation} 
The speed of sound is plotted in Fig. 3. We find that the speed of sound is less than the speed of light.

For the energy conditions we require the quantities
\begin{subequations}
\label{38}
\begin{eqnarray}
\rho - p & = & \frac{6 (1 - a)}{x} - \frac{6 a (x - 4 \alpha ( a - 1)) K (x)}{x^{2}}, \label{38a} \\
\rho + p & = & \frac{6 a (x - 4 \alpha ( a - 1)) K (x)}{x^{2}}, \label{38b} \\
\rho + 3 p & = & \frac{6 (a - 1)}{x} + \frac{18 a (x - 4 \alpha ( a - 1)) K (x)}{x^{2}}. \label{38c}
\end{eqnarray}
\end{subequations}
They are plotted in Fig. 4. The energy conditions $ \rho - p > 0 $, $ \rho + p > 0 $ and $ \rho + 3 p > 0 $ are satisfied in the interior. We note that, close to the centre, the weak energy condition $ \rho - p $ has different behaviour.

We observe that with the help of (\ref{35}) and (\ref{36}) we can write 
\begin{equation}
p = \frac{2 a \rho (3 + 4 \alpha \rho)}{3 (1 - a)} K(\rho) - \rho, \label{39}
\end{equation}
where
\begin{equation}
K(\rho) =  15 \left[ \frac{8\rho ^{2} A^{2} - 4 A (a - 1)(6 + E ) \rho + (a - 1) ^{2}(6 + E) (12 + E)  }{40 \rho ^{2} A^{2} - 15 A (6 + E) \rho + 3 (a - 1) ^{2} (6 + E) (12 + E) } \right].
\end{equation}
Hence the barotropic equation of state (\ref{28}) can be written explicitly for this example. If $ \alpha = 0 $ then the equation of state is
\begin{equation}
p = \left( \frac{10 a}{(1 - a)} - 1 \right) \rho, \label{42}
\end{equation}
which is the Einstein limit. We have generated the linear equation of state (cf. with the results of Case C in Section VI).

\begin{figure}[h!]
  \includegraphics[width=8cm]{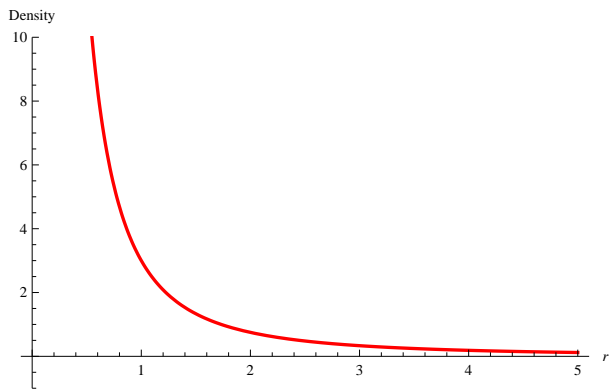}\\
  \caption{Plot of energy density versus radial coordinate $r$}\label{1}
\end{figure}

\begin{figure}[h!]
  \includegraphics[width=8cm]{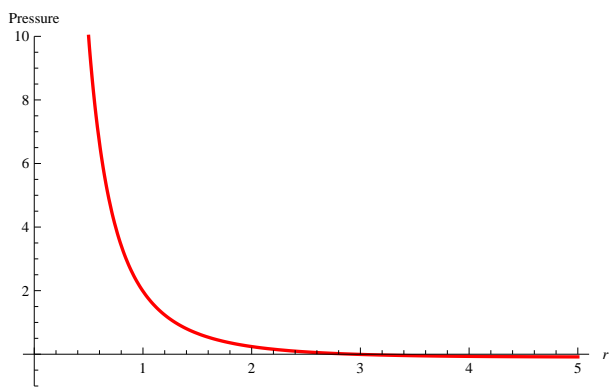}\\
  \caption{Plot of pressure versus radial coordinate $r$}\label{2}
\end{figure}
\newpage
\begin{figure}[h!]
  \includegraphics[width=8cm]{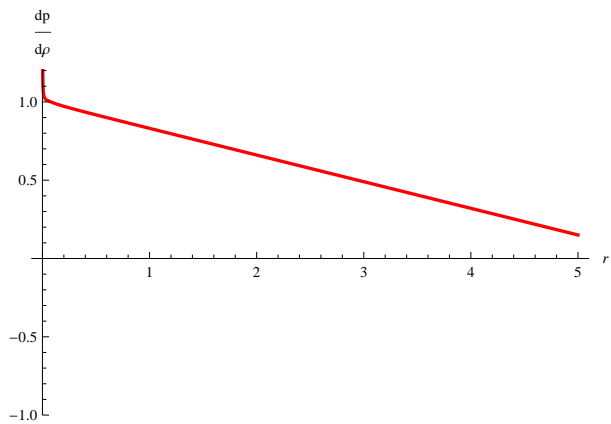}\\
  \caption{Plot of sound-speed parameter versus radial coordinate $r$}\label{3}
\end{figure}

\begin{figure}[h!]
  \includegraphics[width=8cm]{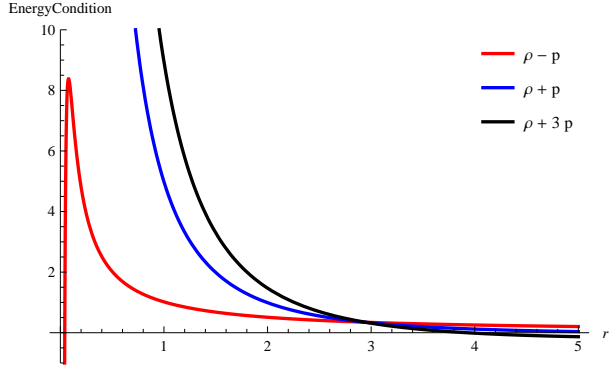}\\
  \caption{Plot of energy conditions versus radial coordinate $r$}\label{4}
\end{figure}

\section{Conclusion}

We have have obtained an interior exact solution for a spherically symmetric sphere in EGB theory coupled with the Lanczos term. The first gravitational potential $ Z $ is taken to be constant and used to find the structure of the second potential $ y $. The method of Frobenius was used to find the first solution in terms of a convergent series. For particular values it is possible to find solutions expressed as polynomials. The second solution is also in the form of a series. The EGB model found always admits a barotropic equation of state. We find that, in both the EGB and Einstein cases, models exist with a linear equation of state with a energy density profile consistent with an isothermal distribution. For the physical analysis we choose a quintic form for the function $ y(x) $. The model is well behaved in the interior as illustrated in graphical plots. The simple models presented here suggest that there may be other exact solutions to the EGB nonlinear equations which we should attempt to discover.

On physical grounds it would be desirable to build the stellar model with an equation of state in the form $ p = p(\rho) $. However note that this requirement adds another nonlinear constraint to the model in addition to the field equations
(\ref{9a})--(\ref{9c}). With this addition we cannot integrate the field equations in general. Even with vanishing EGB coupling parameter $ \alpha $ very few solutions are known in the Einstein case \cite{28} for isotropic pressures. It is therefore remarkable that the class of models found in this paper, with the mathematical assumption (\ref{11}) for the metric, does admit an equation of state. For anisotropic pressures and nonvanishing electric fields there is greater freedom and recently several exact models have been found with an equation of state. Mafa Takisa et al \cite{29}, Maharaj et al \cite{21} presented models with linear equations of state, Maharaj and Mafa Takisa \cite{30} and Mafa Takisa et al \cite{31} generated solutions with a quadratic equation of state, and polytropic equations of state were given by Mafa Takisa and Maharaj \cite{32}. It would be interesting to pursue this approach in the future in EGB gravity.

\section*{ACKNOWLEDGEMENTS}
\noindent We are grateful to the referee for insightful comments that have improved the manuscript. BC thanks the University of KwaZulu-Natal for a scholarship. BC and SH thank the National Research Foundation for financial support. SDM acknowledges that this work is based upon research supported by the South African Research Chair Initiative of the Department of Science and Technology and the National Research Foundation.
 
\bibliography{basename of .bib file}

\end{document}